\documentclass[showpacs,aps,pre,amsmath,amssymb,twocolumn,notitlepage,superscriptaddress,a4paper]{revtex4-1}

\usepackage{graphicx}% Include figure files
\usepackage{dcolumn}% Align table columns on decimal point
\usepackage{bm}% bold math
\usepackage{hyperref}% add hypertext capabilities
\usepackage{color}
\usepackage{natbib}
\begin{document}

%\preprint{}

\title{Crazy 1d billiards: \\ Behavior of  spring-fixated, noisy  colliding particles}

\author{Roman Mani}
\email[Electronic address: ]{manir@ethz.ch}
\affiliation{Computational Physics, IfB, ETH Zurich, Wolfgang-Pauli-Strasse 27, 8093 Zurich, Switzerland}
\author{Lucas B\"ottcher}
\affiliation{ETH Zurich, Clausiusstrasse 50, 8092 Zurich, Switzerland}
\author{Hans J. Herrmann}
\affiliation{Computational Physics, IfB, ETH Zurich, Wolfgang-Pauli-Strasse 27, 8093 Zurich, Switzerland}
\author{Dirk Helbing}
\affiliation{ETH Zurich, Clausiusstrasse 50, 8092 Zurich, Switzerland}
%\thanks{}

%\date{\today}

\begin{abstract}
We study a one-dimensional system of spatially extended particles, which are fixated to regularly spaced locations by means
of elastic springs. The particles are assumed to be driven by a Gaussian noise and to have dissipative, energy-conserving or anti-dissipative
(flipper-like) interactions, when the particle density exceeds a critical threshold. While each particle in separation shows a well-behaved
behavior characterized by a Gaussian velocity distribution, the interaction of particles at high densities can cause an avalanche-like momentum 
and energy transfer, which can generate steep power laws without a
well-defined variance and mean value. Specifically, the velocity variance increases dramatically
towards the free boundaries of the driven-many-particle system. The model might also have some relevance
for a better understanding of crowd disasters. Our results suggest that these are most likely caused by passive momentum transfers, and not by active pushing.
\end{abstract}

\maketitle

\section{Introduction}
In the past, driven many-particle systems and granular media have found a large and continued interest in statistical physics.
For example, in granular systems, one has found the self-organization of collective patterns of motion such as
oscillons \cite{Oscillons}, particle size segregation \cite{Brazil_nut} or the formation of sand dunes, ripples and sheets \cite{Kroy2002,Nishimori1993,Livingstone2007239}.
In traffic flows, scientists have studied the spontaneous emergence of stop-and-go traffic \cite{SugiyamaNJP} and other kinds of congestion patterns \cite{HelbingEPJBPhasediagram,Nagel,Schadschneider,Nagatani}.
In colloidal flows, one has discovered directional segregation phenomena as well as stripe formation in crossing flows \cite{Loewen_etal}.

In pedestrian flows, one has observed the formation of lanes of uniform walking direction in counterflows \cite{MolnarHelbingPRE}, oscillatory flows at bottlenecks \cite{MolnarHelbingPRE},
stop-and-go flows at high densities, and crowd turbulence at even higher ones \cite{HelbingCrowdTurbulencePRE}.
Crowd turbulence has been identified as a common reason of crowd disasters. The phenomenon is characterized by the fact that pedestrians are pushed with a variable and unpredictable intensity and direction,
which is related with certain kinds of power laws. One of the questions often raised is whether the pushing during crowd turbulence is intentional (active) or unintentional (passive). This question is relevant to assess the responsibility
of people for the occurrence of crowd disasters, in which many individuals may die. Here, we discuss a largely abstracted driven many-particle model, which is loosely inspired by this question, but not
claimed to be a model of pedestrian crowds.

We study a number of massive particles of finite radius, which are fixated to regularly spaced locations by elastic springs and can move in one dimension (in a frictionless pipe). The fixation might be considered
to model a ``preferred location'', but this does not matter for our further discussion, as we are dealing here with a theoretically well-defined problem rather than a model of a real system.
If the particles come close enough to each other, they may collide. In the following, we assume hard-core interactions, i.e.~when the distance of two particles with
radius $R$ becomes $2R$, there is an immediate momentum and energy transfer. We will distinguish three different cases: (1) a dissipative case, where energy is absorbed
by collisions (say, transformed into heat), (2) a conservative case, where the kinetic energy stays the same, and (3) an anti-dissipative case, in which the amount of kinetic energy is increased. (The latter case
might be called ``flipper-like'' and could result from intentional pushing, as it would happen if pedestrians wanted to gain space in crowded conditions, as assumed in Refs. \cite{YuJohanssonPRE,PatientImpatient}.)

When collisions do not take place, our model assumes particles to show a linearly damped elastic oscillation, which is driven by a Gaussian noise. As a consequence, the particles will display
normally distributed speeds with identical finite variance, when no collisions take place, i.e. when their distances are sufficiently large. We are interested to find out, how the dynamics of particles changes 
at higher densities, where collisions cause momentum transfers. As we will show, this little modification (i.e. the occurrence of collisions when the average particle distance is reduced) leads to avalanche effects 
which can generate steep power laws, which do not even have a well-defined variance and mean value. Compared to the particles in the bulk of the system, particles tend to have extreme velocity variations towards the free boundaries of the system, if the collisions are dissipative. 
If the collisions are anti-dissipative, the largest velocity variations are observed in the center of the system if the spring damping is sufficiently large. (When comparing this with video recordings of crowd disasters \cite{HelbingCrowdTurbulencePRE,HelbingMukerjiEPJDataScience}, this speaks for passive, i.e. dissipative, 
rather than active, i.e. anti-dissipative, interactions.) In an anti-dissipative system, we even find a finite time singularity, i.e. more or less a diverging dynamics. As a consequence, we can state
that a system of many harmlessly behaving individual components (here: particles with well-defined normally distributed speeds and locations) may show interesting characteristics, when the system elements interact
with each other frequently, as it happens at high densities. This illustrates the surprising dynamics in systems with many well-behaved but strongly interacting system components, as it was 
discussed in Ref. \cite{GloballyNetworkedRisks_inNature}.

\section{Model}
\begin{figure}[htbp]
\begin{center}
\includegraphics[width=0.5\textwidth] {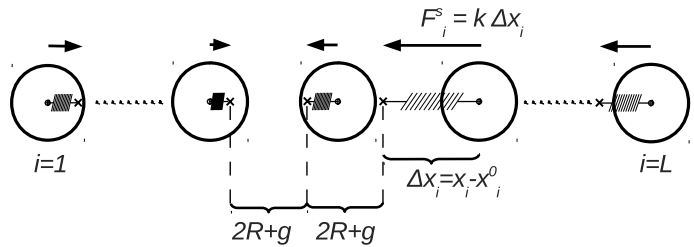}
\caption{Sketch of the model. Arrows indicate the force exerted by the springs
  onto the particles. Crosses denote the spring fixation points.}
\label{fig:sketch}
\end{center}
\end{figure}

We study an assembly of spatially extended particles arranged on a line
where each particle $i$ is attached to a damped spring 
whose origin in $x^0_i$ is fixed. Similar arrangements have been studied before,
though in a different context, for example in Refs. \cite{Hascoet1999,Hascoet2000}. 
The particles are assumed to be driven by
Gaussian noise.
The distance $x_{i+1}^0-x_i^0$ between subsequent spring origins 
is twice the particle radius plus a gap $g:$
\begin{equation}
g=x_{i+1}^0-x_i^0-2R
\end{equation}
A sketch of the assembly
is shown in Fig. \ref{fig:sketch}.
The equation of motion for each particle is:
\begin {equation}
\label{eq:model}
m\ddot x_i = -\gamma \dot x_i -k \Delta x_i + A\xi_i(t),
\end{equation}
where $m$ is the particle mass, $\gamma$ the damping constant of the spring, $k$ the spring constant, $\Delta x_i = x_i-x_i^0$ the distance between the particle position $x_i$ and the spring origin~$x_i^0$, and $\xi_i(t)$ a Gaussian noise with standard deviation~1 and zero mean. The parameter $A$ controls the width of the Gaussian noise and is a measure for the excitation strength of the particles. For a single particle with $k=\gamma=0$ the particle velocity is given by a Wiener Process. 

We employ the Contact Dynamics method \cite{Radjai,brendel_cd,Moreau94} to model perfectly rigid particles.
Hard-core interactions between two particles having velocities $v_1$ respectively $v_2$ are modeled by calculating an interaction force during particle contacts such that the following equation is satisfied:
\begin{equation}
\label{eq:rest}
v_1^{\prime}- v_2^{\prime}=-e(v_1-v_2).
\end{equation}  
Here $v_1- v_2$ and $v_1^{\prime}- v_2^{\prime}$ is the relative velocity of the two particles before, respectively after the collision and $e$ is the particle restitution coefficient. Note that the evolution of the particles due to a contact force automatically ensures momentum conservation in a collision. 

The energy loss or gain $\Delta E$ in a collision is then related to the restitution coefficient $e$ as follows: 
For equal masses ($m=1$ in the following), momentum conservation is given by
\begin{equation}
\label{discrete}
v_1^{\prime}+v_2^{\prime}=v_1+v_2,
\end{equation}  
which is satisfied only if $v_1^{\prime}~=~v_1~+\Delta p$ and $v_2^{\prime}~=~v_2~-~\Delta p$ with the momentum transfer $\Delta p$. 
The energy gain or loss is given by
\begin{equation}
2 \Delta E=(v_1+\Delta p)^2+(v_2-\Delta p)^2-({v_1}^2+{v_2}^2),
\end{equation}  
which, after some algebra, translates to 
\begin{equation}
\label{eq:energy}
2 \Delta E=(e^2-1)(v_1-v_2)^2/2.
\end{equation}  
For $e\leqslant0<1$ we have dissipative collisions, whereas for $e=1$ the energy is conserved. For $e>1$ energy is pumped into the system as it would be the case for a flipper or actively pushing pedestrians.

In the following, we consider particles of equal radii $R$ arranged on a line and we will use $R$ as
our unit of length. The system size $L$ is determined by the
number of particles. The particle mass is set to $m=4\pi/3M$, where  $M$ is the unit of mass and time is measured in units of $T=\sqrt{M/k}$. The time integration of Eq. \eqref{eq:model} is performed by a simple Euler integration 
\begin{eqnarray}
\label{eq:discrete}
v_i(t+\Delta t)&=&v_i(t)+\frac{\Delta t}{m} (-k\Delta x_i-\gamma v_i +A\xi_i/\sqrt{\Delta t}), \nonumber \\
x_i(t+\Delta t)&=&x_i(t )+\Delta t v_i.
\end{eqnarray} 
At the beginning of our computer simulations, the particles
are placed such that $\Delta x=0$ for all particles. For $e\leqslant 1$, the noise drives the particles until the dissipative energy loss due to the damping of the springs and due to collisions equals the energy gain due to noise. We call this state of the system the steady state. All measurements presented in the results section are obtained while the system is in steady state.

\section {Results}
\subsection{Fully dissipative case $e=0$}
\label{seq:e_0}
In this section, we study the fully dissipative case with restitution coefficient $e=0$. In this case, two colliding particles have the same velocities after their collision, but then noise separates them again. The remaining parameters are $k~=~1, A~=~1, \gamma~=~1$ and $g~=~0$.
We first examine the velocity distributions of individual particles in the chain.

\begin{figure}[htbp]
\begin{center}
\includegraphics[width=0.5\textwidth] {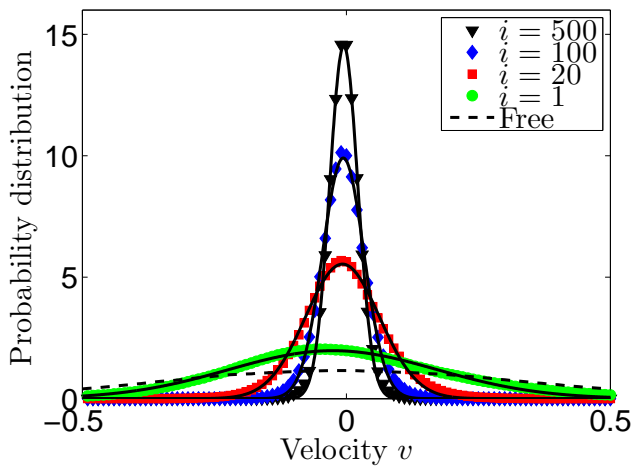}
\caption{(Color online) Velocity distributions of the outermost particle $i=1$, the particle $i=500$
  in the center of the system  and two more particles in between for
  $e=0$, $L=1000$, $\gamma=1$, $A=1$ and $k=1$. Solid lines are Gaussian fits to the data and the dashed line
  corresponds to the free case.}
\label{fig.1}
\end{center}
\end{figure}

\begin{figure}[htbp]
\begin{center}
\includegraphics[width=0.5\textwidth] {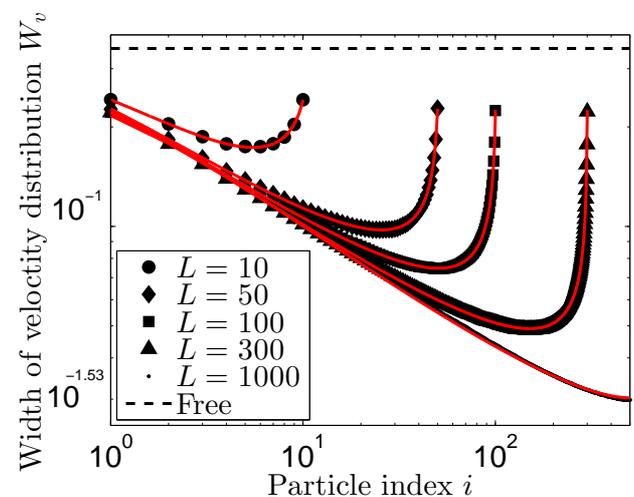}
\caption{(Color online) Width of the velocity distribution $W_v$ versus the particle number $i$ for
  different system sizes $L$ and $e=0$ , $\gamma=1$, $A=1$ and $k=1$. Solid lines
  are power-law fits to the data according to Eq.~\eqref{eq:power_law} and the dashed
  line corresponds to the free case. {Note that the
    log-scale on the  axis distorts the curve, indeed the data is
  reflection-symmetric with respect to $i=(L+1)/2$.}}
\label{fig.2}
\end{center}
\end{figure}

\begin{figure}[htbp]
\begin{center}
\includegraphics[width=0.5\textwidth] {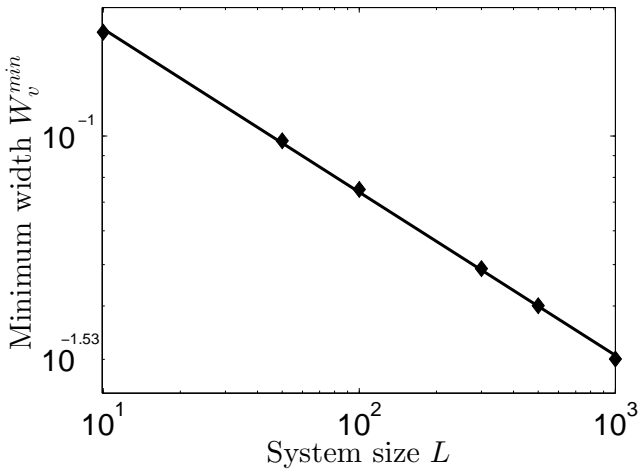}
\caption{Minimum width $W_v^{min}$ as a function of system size $L$ for $e=0$,
  $\gamma=1$, $A=1$ and $k=1$. The line is a power law fit with exponent -0.38.}
\label{fig.3}
\end{center}
\end{figure}

Fig.~\ref{fig.1} shows the velocity distributions measured in steady state for different particles $i$ for a system of size $L=1000$, where $i=1$ corresponds to the particle at the left border and $i=500$ is a particle in the center of the system. In the bulk of the system, the distributions can be fitted by Gaussians. Interestingly, we find that the width of the distributions decreases as a function of the distance to the border of the system. 
For particles close to the border of the system, the probability density functions are slightly skewed. Denoting the probability distribution function by $f_v$, we define the width of the velocity distribution as
\begin{equation}
W_v(i)=\sqrt{\vphantom{\sum_k^a} \int f_v(i)v^2 dv} \sim
\sqrt{\vphantom{\sum_k^a} \sum_k (v^k_i)^2/C },
\end{equation}
where the sum runs over all measurements (at different times in steady state) and is normalized by the total number of measurements $C$.
Note that, since the mass is the same for all particles, the square width $W_v(i)^2$ is proportional to the kinetic
energy of the particle. We want to relate $W_v$ to the position of the particles inside the chain and 
 to the system size. The exact solution of the equation of motion \eqref{eq:model} for the free case, where no collisions occur,  can be found in Ref. \cite{Risken} and the width of the Gaussian velocity distribution is 
\begin{equation}
\label{free}
W_v=\frac{A}{\sqrt{2m\gamma}}.
\end{equation}
In Fig. \ref{fig.2}, we show $W_v(i)$ for different system sizes together with the free case solution Eq. \eqref{free} (dashed line). {Since computation time in Contact Dynamics scales as $L^2$ \cite{brendel_cd}
and the accurate computation of $W_v$ requires a long simulation duration, 
we were limited to system sizes of $L=1000$.} The width of the momentum distribution $W_v$ is decreased compared to the free case solution due to collisions. 
The functions can be fitted by a sum of two power laws as follows
\begin{equation}
\label{eq:power_law}
W_v(i)\sim a\left [(i-c)^{\nu}+(L+1-c-i)^{\nu} \right ]\quad i\in [1..L]
\end{equation}
Note that this form is reflection-symmetric around the center of the system at
$i=(L+1)/2$. Within the error bars, the exponents $\nu\sim -0.46$ are
identical for all system sizes. Note that due to $-1\leqslant-\nu\leqslant 0$ this kind of power-law has no
well-defined mean nor variance \cite{Mendes}. 
Also the  width $W_v^{min}(i=(L+1)/2)$ of the velocity distribution of the central particle follows a power law as a function of the system size $L$ with exponent -0.38 as can be seen in Fig. \ref{fig.3}.

\begin{figure}[htbp]
\begin{center}
\includegraphics[width=0.5\textwidth] {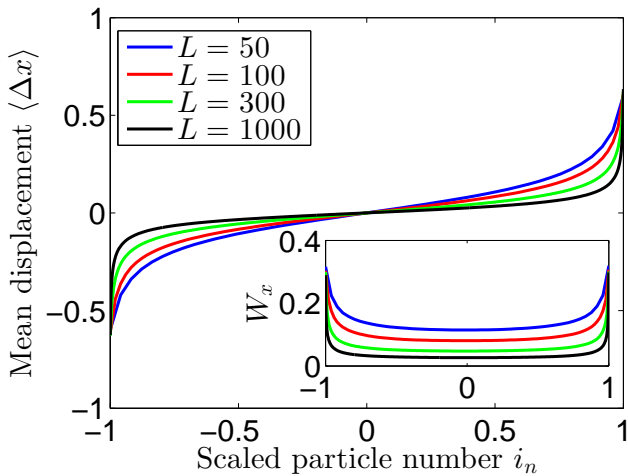}
\caption{(Color online) Mean displacement $\langle \Delta x \rangle$ versus particle number
  $i_n$ for different system sizes and model parameters $e=0$,
  $\gamma=1$, $A=1$ and $k=1$. Inset: Position variance $W_x$ versus scaled particle number $i_n$.}
\label{fig.3_2}
\end{center}
\end{figure}

\begin{figure}[htbp]
\begin{center}
\includegraphics[width=0.5\textwidth] {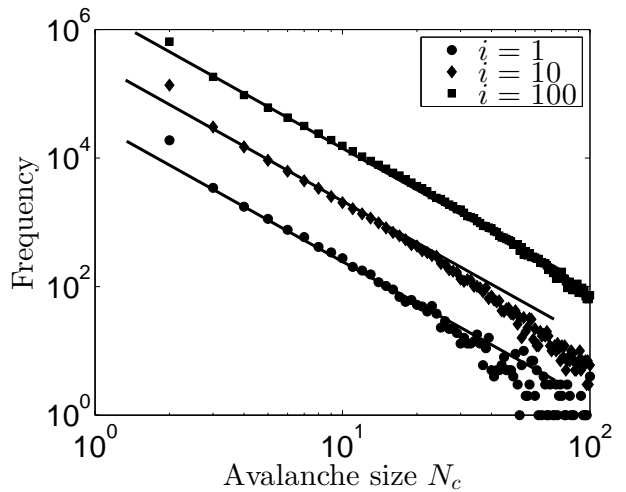}
\caption{Size distribution of triggered collisions  for $e=0$,
  $\gamma=1$, $A=1$, $k=1$ and system size $L=1000$. Frequency is referred to as being the total
  number of occurrences of an avalanche of size $N_c$ during the simulation.}
\label{fig:trigger}
\end{center}
\end{figure}

Next, we analyze the mean displacements $\langle \Delta x\rangle$ for different system sizes. The displacement determines
 the average force exerted by the springs onto the particles.
In order to compare the displacements $\langle \Delta x\rangle$ for different system sizes we define a normalized particle
number $i_n=\frac{2i-(L+1)}{(L-1)}$ such that $i_n(1)=-1$ and $i_n(L)=1$. In Fig. \ref{fig.3_2} we show $\langle \Delta x \rangle$
as a function of $i_n$. Increasing the system size leads to decreasing displacements and less space available for the
particles to move. The more particles are contained in the system, the larger
is the pressure in its interior. To get a better feeling at each position, we calculate the scatter defined as $W_x=\sqrt{\langle \Delta x ^2\rangle -\langle \Delta x\rangle^2}$, which is shown in the inset of Fig. \ref{fig.3_2}. For increasing system sizes, the particles in the interior are  pushed together more and more resulting in a stronger confinement of the particles. 

The power law distributions of $W_v$ suggest cascading effects: Momentum can be transferred via collisions over large length scales. For this
purpose, we measure the distribution of the number $N_c$ of triggered collisions,
where $N_c$ is defined  as follows: Whenever a particular particle $i$ hits
the right neighbor $i+1$, we track how far its momentum is transferred into
the system. If particle $i+1$ just returns without colliding with particle
$i+2$, the momentum is not transferred further. Also, if particle $i+1$
collides with particle $i+2$ and the sum of the velocities (resp. momenta) of
the particle pair is negative, the momentum transfer is interrupted. Only if
the total momentum of the colliding particles is positive (pointing into the
direction of momentum propagation) the transfer continues to particle $i+2$. The number of such triggered
subsequent collisions is called $N_c$. 
Fig. \ref{fig:trigger} shows the distribution of $N_c$ for momentum transfer
waves starting at different particles $i$. The straight lines in Fig. \ref{fig:trigger} have the same
slopes and are guides to the eye, suggesting that the size distribution
follows a power law with exponent around $-2.2$. For non-zero values of the
spring fixation point separation $g$, the large length scale correlations
break down and the bulk of the system becomes insensitive to boundary
effects (see the Appendix for more details).

\subsection{Influence of the particle restitution coefficients}
\label{seq:e_larger}
\begin{figure}[htbp]
\begin{center}
\includegraphics[width=0.5\textwidth] {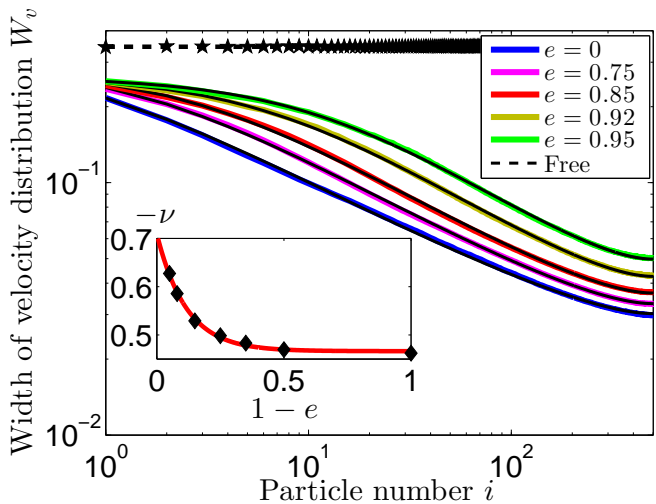}
\caption{(Color online) Velocity variance $W_v$ as a function of the particle number $i$ for
  different values of~$e$ and model parameters $\gamma=1$, $A=1$, $k=1$, $L=1000$. The black lines are power-law fits according to
Eq. \eqref{eq:power_law}.  We only plot the left
  half of the system up to $i=500$. Again, the data is reflection-symmetric with respect to the
  center of the system. The dashed line corresponds to the free case and the stars show $W_v$ for $e=1$ and $L=100$. Inset: Fit of an exponential function to the exponent $-\nu$ in dependence of $1-e$.}
\label{fig.4}
\end{center}
\end{figure}

A similar analysis can be performed for different restitution coefficients: Fig. \ref{fig.4} shows the width of the velocity distribution $W_v$ for different values of the particle restitution coefficient $e$. Again, we consider the case of high density with $g=0$.

\begin{figure}[htbp]
\begin{center}
\includegraphics[width=0.5\textwidth] {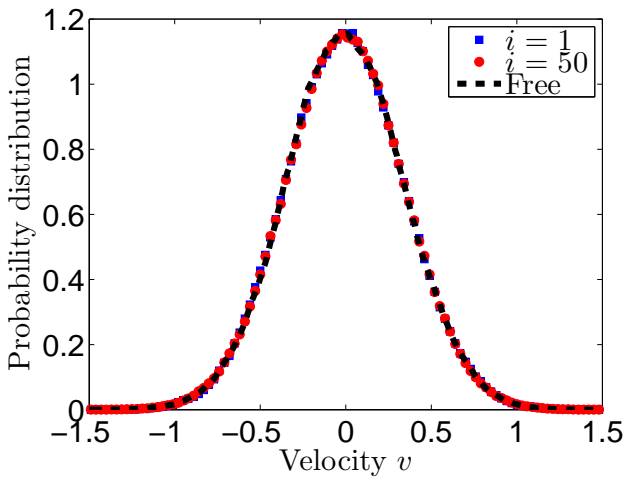}
\caption{(Color online) Velocity distributions of the boundary particle and a particle in
  the center of the system for the non-dissipative case $e=1$, $\gamma=1$, $A=1$, $k=1$ and $L=100$. When interactions are energy conserving, the velocity distributions do not depend on the particle positions. The dashed line is  Gaussian and corresponds to the free case where no collisions occur.}
\label{fig.4.2}
\end{center}
\end{figure}
\begin{figure}[htbp]
\begin{center}
\includegraphics[width=0.5\textwidth] {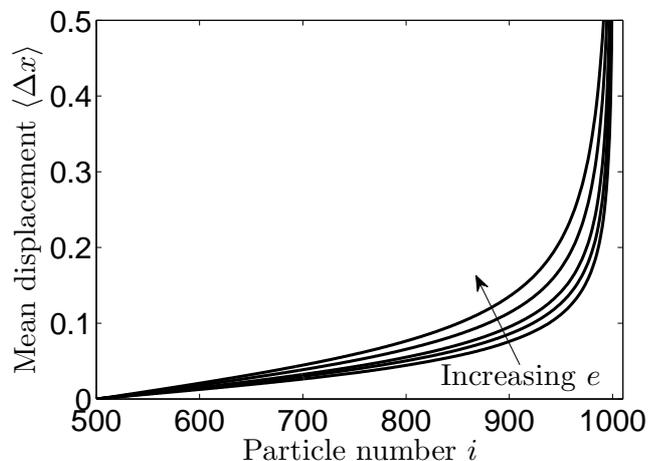}
\caption{The mean displacement $\langle \Delta x\rangle =\langle x-x_0 \rangle
  $ as a function of particle number $i$ for different values of $e~=~(0, 0.75,0.85,0.92,0.95)$ and model parameters
  $\gamma=1$, $A=1$, $k=1$, $L=1000$.}
\label{fig.5}
\end{center}
\end{figure}

The variations of the width and thus kinetic energy become less pronounced for
larger values of $e$. For the case of $e=1$, the distributions become even
independent of the particle number $i$ and correspond to the free case,
 as shown in Fig. \ref{fig.4.2} where the velocity
distributions of a particle at the boundary $i=1$ and in the center $i=50$
are shown for  system of size $L=100$. Again, $W_v$ can accurately be
described by a sum of power laws according to Eq. \eqref{eq:power_law}. The
exponents here depend on $e$ as shown in the inset of Fig. \ref{fig.4}. The
data points were fitted to an exponential $-\nu \sim \mbox{e}^{-\lambda
  (1-e)}+d$. For increasing values of $e$, the variations in $W_v$ and thus,
in kinetic energy are more and more suppressed. However, the mean displacement
$\langle\Delta x \rangle$ of the particles becomes more pronounced for larger
values of $e$ (see Fig. \ref{fig.5}), i.e. the system expands more for large
values of $e$, when collisions are little or not dissipative, leaving more space
for the inner particles to move. Note that $\langle \Delta x \rangle$ does not
follow a power law. In Section \ref{seq:analytics}, we show that  $\langle \Delta x \rangle$ can be described by an $\mathrm{arctanh}$-function for $e\to1$. 

Thus, also for large values of $e$, the total energy is increased at the border
of the system due to the increase in potential energy $\sim k\langle
x^2\rangle$, even though the kinetic energy $\sim W_v^2$ becomes evenly
distributed for $e\to1$. Our numerical results can also be compared to
analytical calculations. For $e=1$, we find excellent agreement between
computer simulation and analytics which will be presented in Section \ref{seq:analytics}.

\subsection{Anti-Dissipation}
\begin{figure}[htbp]
\begin{center}
\includegraphics[width=0.5\textwidth] {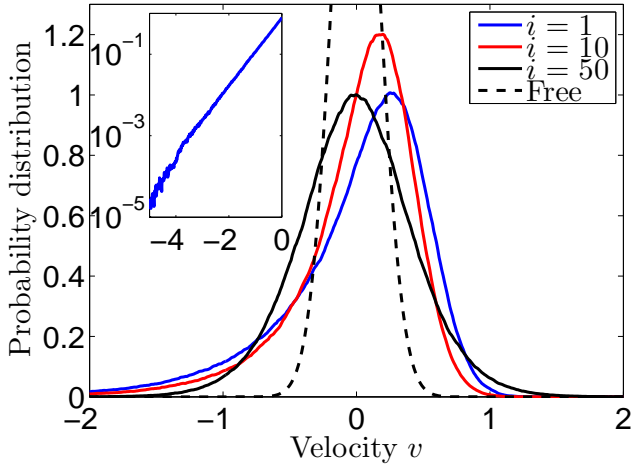}
\caption{(Color online) Velocity distributions for an anti-dissipative system of size
  $L=100$ with model parameters $A=1$, $\gamma=4$ and $e=1.05$. We show distributions for a particle at
  the boundary $i=1$, for a particle in the center $i=50$ and one additional
  particle in between $i=10$. Inset: The curve for $i=1$ on a log-scale, showing only
  negative velocities, indicates an exponential tail. The dashed line corresponds to the case where no collisions occur.}
\label{fig:mom_hist_antidiss_gamma4}
\end{center}
\end{figure}

In this section, we study the anti-dissipative case with $e>1$, where the particles gain energy in collisions and the only dissipation arises due to the spring damping $\gamma$. We focus on the case $g=0$. For $\gamma$ large enough, such that the rate of dissipation is larger than the rate of energy gain in collisions, the system reaches a steady state where the energy fluctuates around a constant mean value. For too small values of $\gamma$, the energy diverges. For $\gamma=4$ and $e=1.05$ the system reaches a steady state and the corresponding velocity probability distributions can be seen in Fig. \ref{fig:mom_hist_antidiss_gamma4}. Here, the velocity distribution of  the outermost particle $(i=1)$ develops a wide tail, falling off exponentially. The width of the velocity distributions  for all particles is larger than the width in the free case (dashed line), meaning that energy is pumped into the system.

\begin{figure}[htbp]
\begin{center}
\includegraphics[width=0.5\textwidth] {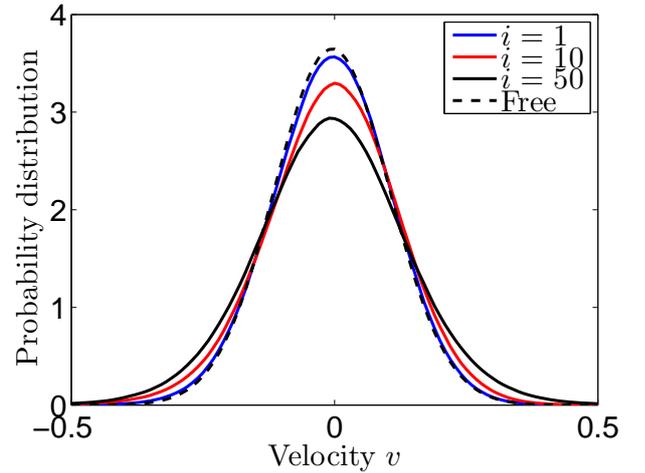}
\caption{(Color online) Velocity distributions for an anti-dissipative system of size $L=100$
  and $A=1$, $\gamma=10$ and $e=1.05$. We show distributions for a particle at the boundary $i=1$, for a particle in the center $i=50$ and one additional particle in between $i=10$. The dashed line corresponds to the case, where no collisions occur.}
\label{fig:mom_hist_antidiss_gamma10}
\end{center}
\end{figure}

The picture changes for larger values of $\gamma$. Fig. \ref{fig:mom_hist_antidiss_gamma10} shows the velocity distributions for the same particles with $\gamma=10$ and $e=1.05$. Interestingly, as opposed to the previous case, where the probability of finding a large negative velocity was larger for the outer particle $(i=1)$ than for the particle in the center of the system, now the probability of finding large velocities is greater for particles in the center of the system. 

\begin{figure}[htbp]
\begin{center}
\includegraphics[width=0.5\textwidth] {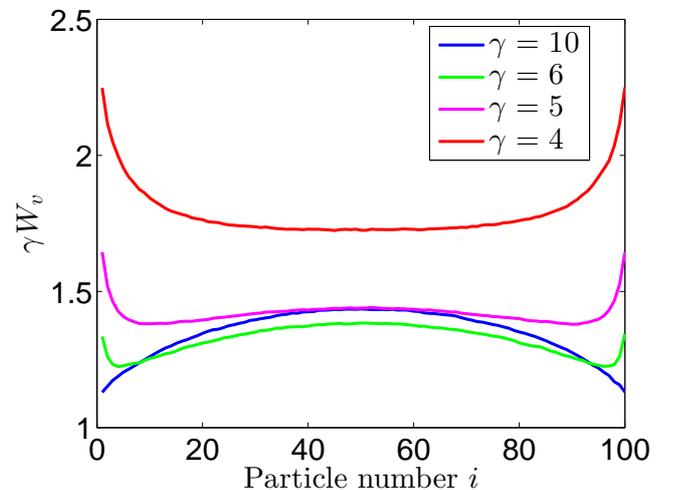}
\caption{(Color online) Width of the velocity distribution $W_v$ multiplied by $\gamma$ as a
  function of the particle number $i$ for $e=1.05$, $A=1$ and different values of the
  spring damping $\gamma$.}
\label{fig:mom_var_antidiss_diff_gamma_sq}
\end{center}
\end{figure}

\begin{figure}[htbp]
\begin{center}
\includegraphics[width=0.5\textwidth] {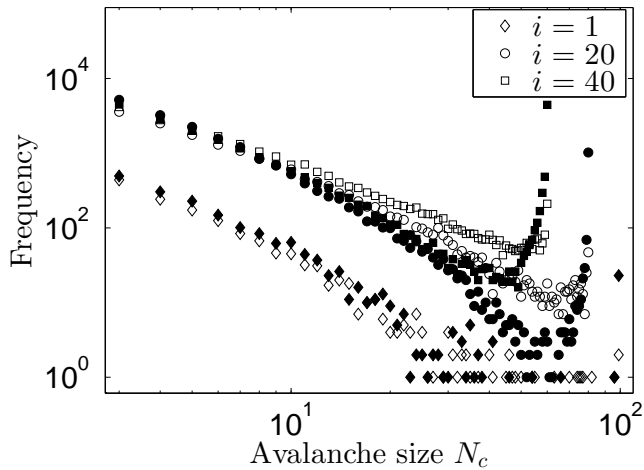}
\caption{ Avalanche size distribution of triggered collisions for $e=1.05$,
  $L=100$ and $A=1$
  triggered by different particles $i$. Frequency is referred to as being the
  total number of occurrences of an avalanche of size $N_c$ during the simulation. Open symbols are for $\gamma=10$ and filled symbols are for $\gamma=4$.}
\label{fig:antidiss_trigger}
\end{center}
\end{figure}

Fig. \ref{fig:mom_var_antidiss_diff_gamma_sq} shows that, as $\gamma$ is increased, the width of the velocity distribution as  a function of the particle number $i$  changes from a convex to a concave shape as $\gamma$ is increased. The concave shape for $\gamma=10$ results from the collision frequency, which is larger in the bulk of the system. Since, for $e>1$, particles gain energy in collisions, the kinetic energy increases in regions of high collision frequency as it is expected for active pushing.  Also for $\gamma=4$ the collision frequency is larger in the center of the system.  But why does the distribution become convex as gamma is decreased? To shed light on this question, we again have a closer look at the avalanche size distributions $N_c$. As can be observed in Fig. \ref{fig:antidiss_trigger}, the probability for finding an avalanche going through the whole system is much larger for a smaller value of the spring damping $\gamma$. In such avalanches the velocity is amplified strongly and is transferred  to the boundary particles such that they gain much kinetic energy. On the other hand, for large values of $\gamma$, the avalanches seldomly go through the whole system and momentum transfer towards the boundary of the system is reduced. Note that the convex distribution for $\gamma=4$ cannot be described by a power-law such as introduced in section \ref{seq:e_0}.

\begin{figure}[htbp]
\begin{center}
\includegraphics[width=0.5\textwidth] {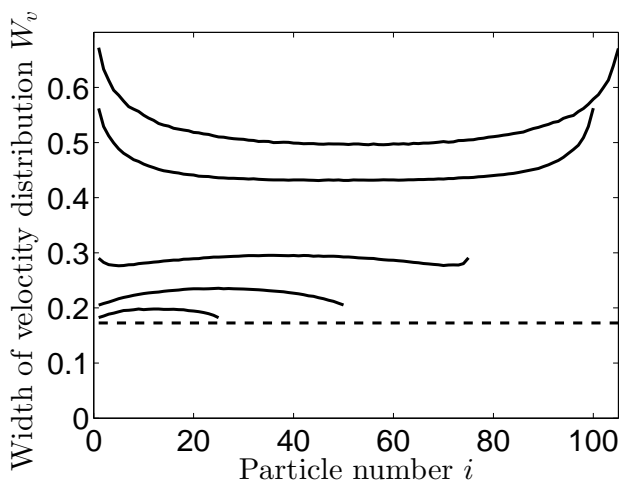}
\caption{Width of the velocity distribution $W_v$ for different system sizes $L=(25, 50, 75, 100, 105)$
  in dependence of the particle number $i$ for $e=1.05$, $A=1$ and $\gamma=4$. The dashed line corresponds to the free case where no collisions occur.}
\label{fig:antidiss_diff_L}
\end{center}
\end{figure}

{For fixed spring damping $\gamma$, also the system size $L$ affects the shape of the width of the velocity distributions~$W_v$:  Fig. \ref{fig:antidiss_diff_L} shows $W_v$ for different system sizes and model parameters $\gamma=4$, $e=1.05$. For small system sizes, $W_v$ is concave, while for larger system sizes, $W_v$ becomes convex. For too large system sizes, the energy diverges. Compared to the free case (dashed line), $W_v$ increases with increasing system size which is due to more collisions occurring. Thus, also more energy is pumped into the system.}

\begin{figure}[htbp]
\begin{center}
\includegraphics[width=0.5\textwidth]{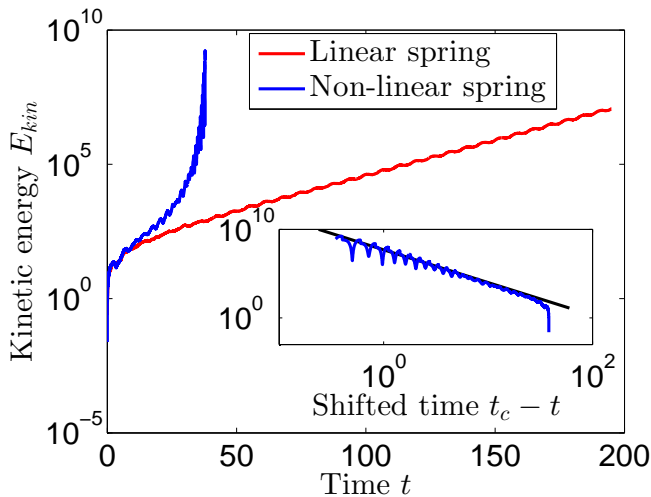}
\caption{(Color online) Kinetic energy $E_{kin}$ as a function of time for $e=1.01$ with
  linear as well as non-linear springs for $\gamma=0.4$, $L=100$, $A=1$ and $k=1$. Inset: Kinetic energy for the non-linear springs as a function of the shifted time $t_c-t$ on a log-log scale.}
\label{non_linear_vs_linear}
\end{center}
\end{figure}

If the spring damping is too small {(respectively, the number of particles is too large)}, i.e. more energy is created than dissipated, the system is constantly gaining energy. For a linear spring, the collision frequency should not depend on the velocity of the particles since the oscillation frequency only depends on the spring damping and spring constant. In each collision, the particles gain an amount of energy proportional to the kinetic energy (see Eq. \eqref{eq:energy}). However, they also lose energy due to the spring damping  proportionally to their kinetic energy, such that the overall kinetic energy is expected to rise as 
\begin {equation}
\frac{dE_{kin}}{dt}\sim C_f E_{kin}-\gamma E_{kin}.
\end{equation}
When the collision frequency $C_f$ is assumed to be approximately constant, this implies an exponential growth of the kinetic energy. This is indeed the case. Fig. \ref{non_linear_vs_linear} demonstrates that the kinetic energy of the system with linear springs rises exponentially for a small enough value of $\gamma$.

{Extreme dynamics, as discussed in Ref. \cite{Bettencourt24042007}, is
characterized by an increase of the kinetic energy faster than
exponential. This may lead to a divergence of the kinetic energy in finite
time, the so called finite time singularity, which is characteristic for catastrophic events.
Such a growth in kinetic energy is for example obtained when
fixating the particles by a non-linear spring, such that the magnitude of the
force is given by $k (x-x_i)^2$.} Then, the oscillation frequency of the spring depends on the particle velocity. The equation of motion for each particle becomes
\begin {equation}
m\ddot x_i = -\gamma \dot {x_i} -k \Delta {x_i}^2\mathrm{sign}(\Delta x_i) + A\xi_i(t).
\end{equation}
For such a system, the collision frequency $C_f$ increases with velocity or kinetic energy like
\begin {equation}
\frac{dE_{kin}}{dt}\sim C_f(E_{kin}) E_{kin} \sim E_{kin}^\beta,
\end{equation}
where $\beta>1$. This implies that the kinetic energy increases faster than exponential, which can be observed in Fig. \ref{non_linear_vs_linear}.
{The inset in Fig. \ref{non_linear_vs_linear} shows the diverging kinetic
energy as a function of $t_c-t$.  $t_c$ denotes the critical time, where the
divergency occurs.  The straight line indicates a power-law behavior near the
critical time $t_c$ and the occurrence of a finite time singularity
\cite{Bettencourt24042007} where the kinetic energy becomes infinite. This
illustrates how the dynamics can turn extreme, even uncontrollable, if
well-behaved but strongly interacting particles are brought together at
high densities.

\section{Comparison to analytical calculations}
\label{seq:analytics}
Here, we explore if it is possible to describe the system of the crazy
billards by a gas-kinetic approach. In Section \ref{seq:e_larger}, 
we saw that for $e=1$, the velocity
distributions of all particles, regardless of their positions, 
can be well described by a Gaussian distribution. Thus, a fluid-dynamic
Maxwell-Boltzmann type equation for compressible gases should apply in this
case. A suitable equation,  where the finite
size of the particles as well as granular collisions are taken into account is given in Ref. \cite{Helbing1996}:
\begin{equation}
\label{eq:helbing}
\frac{\partial\tilde{\rho}}{\partial t} + v \frac{\partial}{\partial x} \tilde{\rho} + \frac{\partial}{\partial v}\left(\tilde{\rho}\frac{d v}{d t}\right) = \frac{1}{2} \frac{\partial^2}{\partial v^2}\left(D\tilde{\rho}\right)+\left(\frac{\partial\tilde{\rho}}{\partial t}\right)_{int}.
\end{equation}
Here,  $\tilde\rho(x,v,t)$ is the phase space density which follows a Gaussian distribution
\begin{equation}
\tilde{\rho} = \frac{\rho}{\sqrt{2\pi}}\exp{\left[-(v-V)^2/2W_v^2)\right]}.
\end{equation}
$V=\langle v \rangle$ is the macroscopic velocity and $\rho$ is the
particle density. The diffusion term containing the diffusion constant $D$ takes into account
the driving of the particles due to the Gaussian noise. The interaction
term $\left({\partial\tilde{\rho}}/{\partial t}\right)_{int}$ (see Appendix) is describing granular collisions with restitution
coefficient $e$.
Multiplication of Eq. \eqref{eq:helbing} by $v$ on both sides, integrating over $v$ and evaluating the interaction term (see Appendix for details) leads to the following equation relating
the macroscopic variables $V$, $\rho$ and $W_v$:
\begin{equation}
\label{modeleq}
\frac {\partial V} {\partial t} + V\frac {\partial V} {\partial x}=-\frac{1}{\rho}\frac{\partial}{\partial x}\frac{\rho W_v^2}{1-2\rho R}-k\langle \Delta x \rangle -\gamma V,
\end{equation}
}
This equation is a slight modification of the equation from Ref. \cite{Helbing1996} to reflect our spring-fixated particles. 
The term containing $W_v^2$ has the meaning of  a pressure.
In the stationary state, $\frac {\partial V} {\partial t}=0$. Moreover, since there is no particle migration in steady state, the macroscopic velocity $V$ vanishes. This leads to
\begin{equation}
\frac{1}{\rho}\frac{\partial}{\partial x}\frac{\rho W_v^2}{1-2\rho R}=-k\langle \Delta x \rangle,
\end{equation}
which can be interpreted as a force balance between the frictional and spring forces (R.H.S) and the force exerted by a pressure gradient (L.H.S). 
Assuming that $W_v^2 \sim const$ as explored in Section \ref{seq:e_larger} for $e=1$, (see Fig. \ref{fig.4.2}) and representing the derivative $\partial \rho/\partial x $ by $\rho'$ we can write
\begin{eqnarray}
\frac{\partial}{\partial x}\frac{\rho W_v^2}{1-2\rho R}&=&W_v^2\frac{\rho'(1-2\rho R)+2\rho \rho'R}{(1-2\rho R)^2}\nonumber\\ &=&W_v^2\frac{\rho'}{(1-2\rho R)^2},
\end{eqnarray}
which leads to 
\begin{equation}
\label{diff_gl}
\frac{\partial \rho}{\partial x}=-\frac{k}{W_v^2}\langle \Delta x \rangle \rho (1-2\rho R)^2.
\end{equation}
For particle masses $m$ different from unity, the right hand side of this equation
must be divided by $m$.
\begin{figure}[htbp]
\begin{center}
\includegraphics[width=0.5\textwidth]{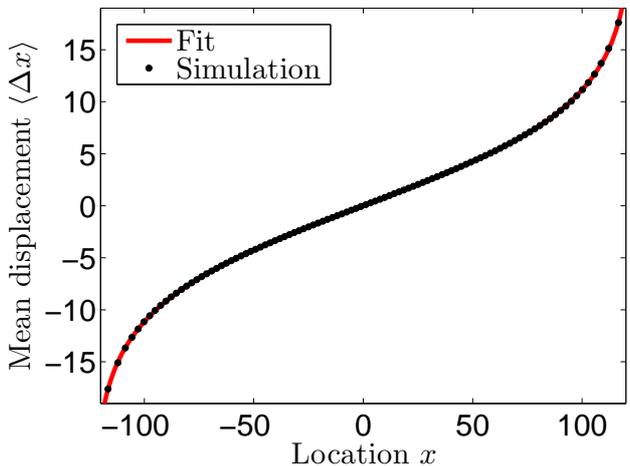}
\caption{(Color online) For $e=1$, the mean displacement $\langle \Delta x \rangle$ (points) can be
  described by a function $f(x)= B\cdot \mathrm{arctanh}(Cx)$ (line) with fit parameters B and C. The model parameters are $A=10$, $\gamma=1$, $k=1$
  and $L=100$.}
\label{fig:atan}
\end{center}
\end{figure}

We consider the case $e=1$ for a chain of $L=100$ particles, having fixation points separated by
$2R$.
Empirically, $\langle \Delta x\rangle$ can be described by a arctanh function as shown in Fig. \ref{fig:atan}. The variable $x$ refers to the real
average position of the particles defined by 
\begin{equation}
x=\{2i-(L+1)\}R+\langle \Delta x_i\rangle.
\end{equation}

\begin{figure}[htbp]
\begin{center}
\includegraphics[width=0.5\textwidth]{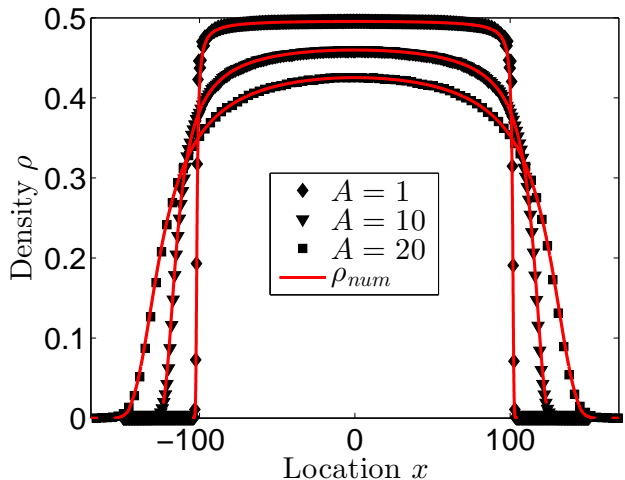}
\caption{(Color online) The particle density as obtained from computer simulations (symbols) and by
  numerical integration of Eq. \eqref{diff_gl} (lines) for different values of
  the driving amplitude $A$. The model parameters are $e=1$, $\gamma=1$, $k=1$, $L=100$. A vanishing density 
  $\rho=0$ (for large $|x|$) means
  that there is no particle found at that particular location during the simulation.
}
\label{fig:rho_num_sim}
\end{center}
\end{figure}

The density $\rho$ can be calculated as follows: The space is divided into
bins of equal size, much smaller than the particle radius. In each
measurement, the values of the bins are incremented by one if the whole bin is
"covered" by a particle. At the edge of a particle, only part of the bin is
covered. Then, the value in the bin is incremented only by the percentage by which
the bin is covered by the particle. In the end, the bin values are divided by
the number of measurements and by $2R$, and finally smoothed, using a running
average filter of window size $2R$. In fact, this is equivalent to convoluting the particle position distribution with a rectangular function $\Theta_H(x-2R)- \Theta_H(x)$ of length 2R, where $\Theta_H$ denotes the Heaviside Theta function. Such a convolution takes into account the finite extent of the particles.  The results for different values of $A$ and $e=1$ are shown in Fig. \ref{fig:rho_num_sim} (symbols). 

We can insert the fitted curve $\Delta x$ together with $W_v^2$ into Eq. \eqref{diff_gl} and perform a numerical integration of Eq. \eqref{diff_gl}, yielding the numerical density $\rho_{num}$ in Fig. \ref{fig:rho_num_sim}. We see that the results match the computer-simulated densities perfectly.  This leads to the conclusion that the system can be described by a gas-kinetic approach.

\section{Conclusion}
In this paper, we studied a system of finite-sized particles, which are driven
by Gaussian noise and fixated by damped springs. At very low densities, where
the particles never collide, the velocity distributions of the particles are
Gaussian, while  at high densities, collisions between neighboring particles
lead to an avalanche-like energy and momentum transfer. For dissipative
collisions, we find that the velocity variance of the particles follows a
power-law towards the free boundary with no well-defined mean and
variance.  The picture
changes when collisions are anti-dissipative. Provided that the spring damping
is large enough, the velocity variations become largest in the bulk of the
system. For strong anti-dissipation, we find an increase of particle energy over time.
For non-linear springs, energy can diverge at a finite time. Finally, we found
good agreement of our numerical results with a gas-kinetic analytical approach.

\begin{widetext}
\section{Appendix}
\subsection{Gas-kinetic approach}
\label{sec:1}
The following derivation of Eq. \eqref{modeleq} is based on a gas-kinetic approach in analogy to  Refs. \cite{Helbing1996} and \cite{verkehrsdynamik}:
\begin{equation}
\label{eq1}
\frac{\partial\tilde{\rho}}{\partial t} + v \frac{\partial}{\partial r} \tilde{\rho} + \frac{\partial}{\partial v}\left(\tilde{\rho}\frac{d v}{d t}\right) = \frac{A^2}{2 m^2} \frac{\partial^2}{\partial v^2}\tilde{\rho}+\left(\frac{\partial\tilde{\rho}}{\partial t}\right)_{int}
\end{equation}
This expression denotes the reduced, gas-kinetic transport equation with an additional diffusion term $A^2 / 2 m^2(\partial^2/\partial v^2)\tilde{\rho}$. We assume that the velocity distribution is a Gaussian:
\begin{equation}
\label{eq2}
\tilde{\rho}(r,v,t) = \frac{\rho(r,t)}{\sqrt{2\pi W_v(r,t)^2}}\exp{\left[-(v-V(r,t))^2/2W_v(r,t)^2\right]}
\end{equation}
The acceleration term for unit mass equals (without any indices $i$):
\begin{equation}
\label{eq3}
\frac{d v}{d t} = -\gamma v - k\langle\Delta x\rangle
\end{equation}
The interaction term can be written as:
%\begin{widetext}
\begin{equation}
\label{eq4}
   \begin{split}
\left(\frac{\partial\tilde{\rho}}{\partial t}\right)_{int} = \int dv^\prime\int dw \int dw^\prime \sigma(v^\prime,w^\prime |v,w)|v^\prime - w^\prime|\tilde{\rho}_2(r,v^{\prime};r+2 R,w^{\prime};t)\\
-\int dv^\prime\int dw \int dw^\prime \sigma(v,w |v^{\prime},w^{\prime})|v - w|\tilde{\rho}_2(r,v;r+2 R,w;t)\\
+\int dv^\prime\int dw \int dw^\prime \sigma(w^\prime,v^\prime |w,v)|v^\prime - w^\prime|\tilde{\rho}_2(r,v^{\prime};r-2 R,w^{\prime};t)\\
-\int dv^\prime\int dw \int dw^\prime \sigma(w,v |w^{\prime},v^{\prime})|v - w|\tilde{\rho}_2(r,v;r-2 R,w;t),
\end{split}
\end{equation}
%\end{widetext}
describing interactions at location $r^{\prime}=r\pm 2 R$. The differential cross section for completely elastic collisions ($e=1$) is given by: 
\begin{equation}
\sigma(v^\prime,w^\prime |v,w)=\delta(v^{\prime}-w)\delta(w^{\prime}-v). 
\end{equation}
The pair distribution function $\tilde{\rho}_2$ can be approximated as:
\begin{equation}
\label{eq5}
\tilde{\rho}_2(r,v;r\pm 2 R,w;t)=\chi(r\pm R,t)\tilde{\rho}(r,v,t)\tilde{\rho}(r\pm 2 R,w,t),
\end{equation}
where the factor $\chi$ is defined as:
\begin{equation}
\label{chi}
\chi(r\pm R,t):=\frac{1}{1-2\rho(x\pm R,t)R},
\end{equation}
denoting the increase in particle interaction, due to the finite extension of $R$ around the center of the oscillating particle. 
We are now integrating Eq. \eqref{eq1} over $v$ from $-\infty$ to $+\infty$, after multiplying it with the collisional invariants $\psi(v)=1$ or $v$. This provides us with the continuity and velocity equation. Using the fundamental theorem of calculus, the rules of interchanging differentiation and integration as well as the identities in section \ref{sec:2} and evaluating the single terms, we find two equations:
\begin{equation}
\label{eq6}
\frac{\partial \rho}{\partial t}+\frac{\partial}{\partial r}\left(\rho V\right)=\int dv \left(\frac{\partial \tilde{\rho}}{\partial t}\right)_{int}
\end{equation}
\begin{equation}
\label{eq7}
\frac{\partial}{\partial t}\left(\rho V \right) +\frac{\partial}{\partial r}\left[\rho(V^2+W_v^2)\right] + \rho\left[\gamma V + k\langle \Delta x\rangle\right]=\int dv v \left(\frac{\partial \tilde{\rho}}{\partial t}\right)_{int}
\end{equation}
In a further step we rewrite the interaction term, in Eq. \eqref{eq4}. Since we are integrating over $v$, it is allowed to interchange $v \leftrightarrow v^{\prime}$ and $w \leftrightarrow w^{\prime}$. Thus, after multiplying Eq. \eqref{eq4} with $\psi(v)$ and integrating over $v$, we find:
%\begin{widetext}
\begin{align}
\label{eq8}
\mathcal{I}(\psi):= & \int dv \int dv^\prime \int dw \int dw^\prime \left[\psi(v^\prime)-\psi(v)\right]\sigma(v,w|v^\prime,w^\prime)|v - w|\notag\\
& \times\tilde{\rho}_2(r,v;r+2 R,w;t)\notag\\
& + \int dv \int dv^\prime \int dw \int dw^\prime \left[\psi(v^\prime)-\psi(v)\right]\sigma(w,v|w^\prime,v^\prime)|v - w|\notag\\
& \times\tilde{\rho}_2(r,v;r-2 R,w;t)
\end{align}
%\end{widetext}
For $\psi(v)=1$ Eq. \eqref{eq8} implies $\mathcal{I}(1)=0$, such that Eq. \eqref{eq6} reduces to a continuity equation:
\begin{equation}
\label{eq9}
\frac{\partial \rho}{\partial t}+\frac{\partial}{\partial r}\left(\rho V\right)=0
\end{equation}
Equation \eqref{eq7} can be reduced to
\begin{equation}
\label{eq10}
\frac{\partial V}{\partial t} +V \frac{\partial V}{\partial r} = -\frac{1}{\rho} \frac{\partial}{\partial r}\left(\rho W_v^2\right) - \left[\gamma V+k\langle \Delta x\rangle\right]+\frac{\mathcal{I}(v)}{\rho}.
\end{equation}
The remaining task is evaluate the interaction term $\mathcal{I}(v)$. We can simplify Eq. \eqref{eq8} again, interchanging $v\leftrightarrow w$ and $v^{\prime}\leftrightarrow w^{\prime}$.
\begin{align}
\label{eq11}
\mathcal{I}(\psi):= & \int dv \int dv^\prime \int dw \int dw^\prime \sigma(v,w|v^\prime,w^\prime)|v - w|\notag\\
& \times\{\left[\psi(v^\prime)-\psi(v)\right]\tilde{\rho}_2(r,v;r+2 R,w;t)\notag\\
& +\left[\psi(w^\prime)-\psi(w)\right]\tilde{\rho}_2(r-2 R,v;r,w;t)\}
\end{align}
Applying a first order Taylor expansion for $\tilde{\rho}_2(r,v;r+2 R,w;t)$ yields:
%\begin{widetext}
\begin{equation}
\label{eq12}
\tilde{\rho}_2(r,v;r+2 R,w;t) = \tilde{\rho}_2(r-2 R,v;r,w;t)+2 R \frac{\partial}{\partial r}\tilde{\rho}_2(r,v;r+2 R,w;t)+\mathcal{O}(R^2)
\end{equation}
%\end{widetext}
We can use Eq. \eqref{eq12} to write the interaction term as
\begin{equation}
\label{eq13}
\mathcal{I}(\psi)=\mathcal{I}_s(\psi)-\frac{\partial \mathcal{I}_f(\psi)}{\partial r}
\end{equation}
with a source term
%\begin{widetext}
\begin{align}
\label{eq14}
\mathcal{I}_s(\psi)=&\int dv \int dv^\prime \int dw \int dw^\prime \sigma(v,w|v^\prime,w^\prime)|v - w|\notag\\
&\times \{  \left[ \psi(v^{\prime})+\psi(w^{\prime})\right]-\left[\psi(v)+\psi(w) \right]  \} \tilde{\rho}_2(r-2 R,v,r,w;t)
\end{align}
%\end{widetext}
and a flux term
\begin{align}
\label{eq15}
\mathcal{I}_f(\psi)=&-2 R\int dv \int dv^\prime \int dw \int dw^\prime \sigma(v,w|v^\prime,w^\prime)|v - w|\notag\\
&\times\left[\psi(v^{\prime})-\psi(v)\right]\tilde{\rho}_2(r,v;r+2 R,w;t)
\end{align}
We see that the source term in Eq. \eqref{eq14} vanishes for $\psi(v)=v$. We proceed with a Taylor expansion of the pair distribution function $\tilde{\rho}_2$:
%\begin{widetext}
\begin{align}
\label{eq16}
\tilde{\rho}_2(r,v;r+2R,w;t)&=\chi(r + R,t)\tilde{\rho}(r,v,t)\tilde{\rho}(r + 2R,w,t) \notag\\
&\approx\chi(r,t)\tilde{\rho}(r,v,t)\tilde{\rho}(r,w,t)\left\{1+\frac{R}{\chi}\frac{\partial\chi}{\partial x}\right.\notag\\
&+2 R\left. \left. \left[\frac{1}{\rho}\frac{\partial \rho}{\partial x}+\frac{(v-V)}{W_v^2}\frac{\partial V}{\partial x}\right.+\frac{1}{2W_v^2}\left(\frac{\left(v-V\right)^2}{W_v^2}-1\right)\frac{\partial W_v^2}{\partial x}\right]\right\}
\end{align}
%\end{widetext}
If we neglect the higher order derivatives in $\partial \mathcal{I}_f(\psi)/\partial r$ and Eq. \eqref{eq16}, there remains only one contribution for $\mathcal{I}_f(v)$:
\begin{equation}
\label{eq17}
\mathcal{I}_f(v)=2 R\chi(x,t)\rho(x,t)^2W_v(x,t)^2.
\end{equation}
For the velocity equation we find, by plugging in Eq. \eqref{eq17} into Eq. \eqref{eq10}:
\begin{equation}
\label{continuity_helbing}
\frac{\partial V}{\partial t} +V \frac{\partial V}{\partial r} = -\frac{1}{\rho} \frac{\partial}{\partial r}\left(\frac{\rho W_v^2}{1-2\rho R}\right) - \left[ \gamma V+k\langle \Delta x\rangle\right]
\end{equation}
\end{widetext}

\subsection{Definitions and Identities}
\label{sec:2}
In a first step we define:
\begin{equation}
P(x,v,t) := \frac{\tilde{\rho}(x,v,t)}{\rho(x,t)}
\end{equation}
For the evaluation of the gas-kinetic equation we have:
\begin{equation}
\frac{\partial\tilde{\rho}}{\partial t} = \frac{\partial}{\partial t}\left(P\rho\right)
\end{equation}
\begin{equation}
v \frac{\partial}{\partial x} \tilde{\rho}= v\frac{\partial}{\partial x}\left(P\rho\right)
\end{equation}
\begin{equation}
\frac{\partial}{\partial v}\left( \tilde{\rho} \frac{d v}{d t}\right)= P\rho \left[ \frac{(v-V)}{W_v^2}(\gamma v + k\Delta x) -\gamma \right]
\end{equation}
\begin{equation}
\frac{A^2}{2 m^2} \frac{\partial^2}{\partial v^2} \tilde{\rho} = \frac{A^2}{2 m^2}P\rho\left [\left( \frac{v-V}{W_v^2} \right)^2-(W_v^2)^{-1}\right]
\end{equation}
Some necessary integrals are:
\begin{equation}
\rho(x,t) = \int_{-\infty}^{\infty} \tilde{\rho}(x,v,t)  dv 
\end{equation}
\begin{equation}
V(x,t) := \langle v \rangle = \int_{-\infty}^{\infty} v P(x,v,t)  dv
\end{equation}
\begin{equation}
W_v^2(x,t) := \langle (v-V)^2 \rangle = \int_{-\infty}^{\infty} (v-V)^2 P(x,v,t)  dv
\end{equation}
These are the evaluated integrals for the calculation:
\begin{align}
& \int_{-\infty}^{\infty} v^2 P(x,v,t)  dv = (V^2+W_v^2) \\
& \int_{-\infty}^{\infty} v (v-V) P(x,v,t)  dv = W_v^2 \\
& \int_{-\infty}^{\infty} v^2 (v-V) P(x,v,t)  dv = 2W_v^2 V \\
& \int_{-\infty}^{\infty} v (v-V)^2 P(x,v,t)  dv = W_v^2 V
\end{align}

\subsection{Influence of driving amplitude $A$ and spring fixation point separation $g$}

\begin{figure}[htbp]
\begin{center}
\includegraphics[width=0.5\textwidth] {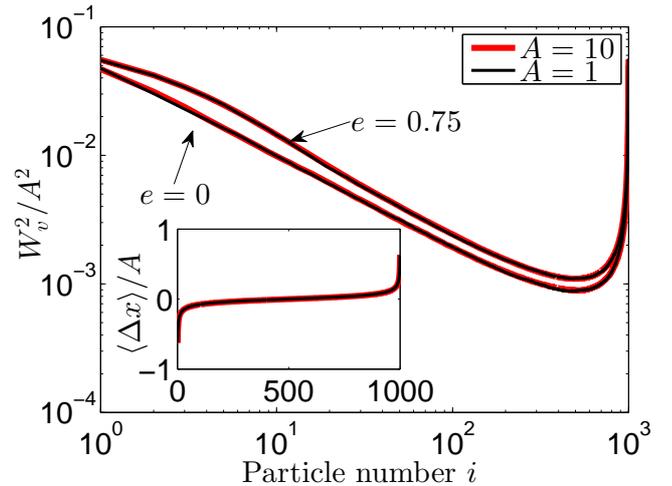}
\caption{(Color online) $\langle W_v^2 \rangle  $ normalized by $A^2$ versus particle number
  for different values of $A$ and $e$ with $\gamma=1$, $k=1$ and $L=1000$. Inset: $\langle \Delta x\rangle  $ normalized by $A$ versus the particle number $i$. }
\label{fig:amp_influence}
\end{center}
\end{figure}

Now we discuss the influence of the model parameters $A$ and $g$. First, we suggest that the results can be rescaled via a power of $A$. For instance $\langle x \rangle /A$, $ W_v^2  /A^2$ and $\langle x^2 \rangle /A^2$ are all constants. We tested this rescaling for  a few cases shown in Fig. \ref{fig:amp_influence}, where $W_v^2$ and $\langle \Delta x \rangle$ for two different values of $e$ collapse onto a single curve. 

Next, we investigate how the system depends on the particle density or spring density. It is plausible that the system goes from a completely uncorrelated state when particles hardly collide at small spring densities towards a completely correlated state, when the particles influence each other over large length scales. 
\begin{figure}[htbp]
\begin{center}
\includegraphics[width=0.5\textwidth] {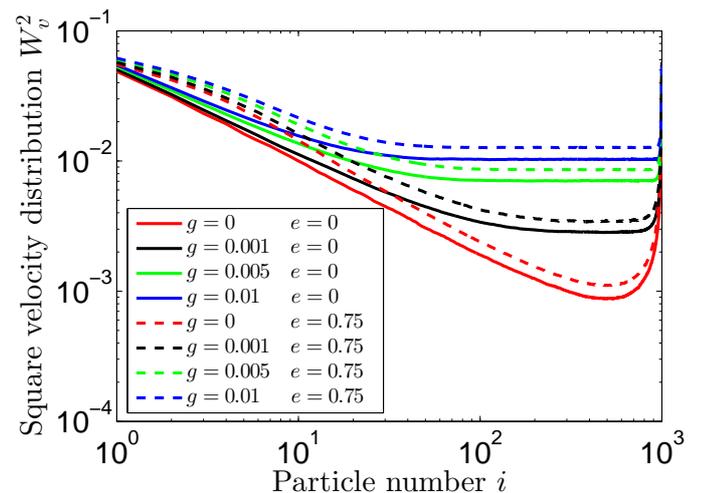}
\caption{(Color online) Square width of the velocity distribution $W_v^2$ versus particle number for different gaps $g$ and different
  values of $e$ for $A=1$, $\gamma=1$, $k=1$ and $L=1000$.}
\label{fig:mom_var_gaps}
\end{center}
\end{figure}
Here we study this influence by
tuning the parameter $g$ defined in Fig. \ref{fig:sketch}. The results are presented in Fig. \ref{fig:mom_var_gaps}, where $W_v^2$ is plotted for different "gaps" $g$. We observe the occurrence of a plateau in the interior of the system, which becomes wider and wider with decreasing spring density. This means that the cascading effects and the spatial correlations are limited in their spatial extension and that there is no sensitivity to boundary effects in the bulk of the system for large system sizes. The particles in the center of the system basically behave as if they move freely, but with an enhanced dissipation due to pairwise collisions, which are still occurring.

{\begin{acknowledgments}
We acknowledge financial support from the  (ERC) Advanced grant number
FP7-319968-FlowCCS of the European Research Council and from the DFG grant
number He 2732/11-3 in the SPP 1486 ``PiKo''. We also acknowledge (partial) support 
by the European Commission through the ERC Advanced Investigator Grant ‘Momentum’ (Grant No. 324247).
\end{acknowledgments}}
\bibliography{science}
\end{document}